%Paper: hep-th/9312108
%From: "Adel Bilal" <bilal@puhep1.Princeton.EDU>
%Date: Mon, 13 Dec 93 14:42:01 -0500

%%%%%%%%%%%%%%%%%%%%%%%%%%%%%%%%%%%%%%%%%%%%%%%%%%%%%%%%%%%%%%%%%%%%%%%%%%

\input phyzzx

%%%%%%%%%%%%%%%%%%%%%%%%%%%%%%%%%%%%%%%%%%%%%%%%%%%%%%%%%%%%%%%%
% This will make your PHYZZX pagesize wider and longer
% It is OPTIONAL. It redefines the papers macro
%
%\catcode`\@=11 % This allows us to modify PLAIN macros.
%
%\def\papers{\papersize\headline=\paperheadline\footline=\paperfootline}
%
%\def\papersize{\hsize=40pc \vsize=53pc \hoffset=0pc \voffset=1pc
%   \advance\hoffset by\HOFFSET \advance\voffset by\VOFFSET
%   \pagebottomfiller=0pc
%   \skip\footins=\bigskipamount \normalspace }
%
%\catcode`\@=12 % at signs are no longer letters
%
%\papers
%%%%%%%%%%%%%%%%%%%%%%%%%%%%%%%%%%%%%%%%%%%%%%%%%%%%%%%%%%%%%%%%

\def\to{\rightarrow}

\vsize=21.5cm
\hsize=15.cm

\tolerance=500000
\overfullrule=0pt

\pubnum={PUPT-1434 \cr
hep-th@xxx/9312108 \cr
December 1993}

\date={}
\pubtype={}
\titlepage
\title{NON ABELIAN TODA THEORY :\break
A COMPLETELY INTEGRABLE MODEL FOR\break
STRINGS ON A BLACK HOLE BACKGROUND}
\author{{
Adel~Bilal}\foot{
 on leave of absence from
Laboratoire de Physique Th\'eorique de l'Ecole
Normale Sup\'erieure, \nextline 24 rue Lhomond, 75231
Paris Cedex 05, France
(unit\'e propre du CNRS)\nextline
e-mail: bilal@puhep1.princeton.edu
}}
\address{\it Joseph Henry Laboratories\break
Princeton University\break
Princeton, NJ
08544, USA}

\vskip 3.mm
\abstract{The present paper studies a completely integrable
conformally invariant model in 1+1 dimensions that corresponds to
string propagation on the two-dimensional black hole background
(semi-ininite cigar). Besides the two space-time string fields there
is a third (internal) field with a very specific Liouville-type
interaction leading to the complete integrability. This system is
known as non-abelian Toda theory. I give the general explicit classical
solution. It realizes a rather involved transformation
expressing the interacting  string fields in terms of (three) functions
$\varphi_j(u)$ and $\bar\varphi_j(v)$ of one light-cone variable only. The
latter are shown to lead to standard harmonic oscillator (free field) Poisson
brackets thus paving the way towards quantization. There are three
left-moving and three right-moving conserved quantities.  The right
(left)-moving conserved quantities form a new closed non-linear, non-local
Poisson bracket algebra. This algebra is a Virasoro algebra extended by two
conformal dimension-two primaries. }

\endpage
\pagenumber=1

 \def\PL #1 #2 #3 {Phys.~Lett.~{\bf #1} (#2) #3}
 \def\NP #1 #2 #3 {Nucl.~Phys.~{\bf #1} (#2) #3}
 \def\PR #1 #2 #3 {Phys.~Rev.~{\bf #1} (#2) #3}
 \def\PRL #1 #2 #3 {Phys.~Rev.~Lett.~{\bf #1} (#2) #3}
 \def\CMP #1 #2 #3 {Comm.~Math.~Phys.~{\bf #1} (#2) #3}
 \def\IJMP #1 #2 #3 {Int.~J.~Mod.~Phys.~{\bf #1} (#2) #3}
 \def\JETP #1 #2 #3 {Sov.~Phys.~JETP.~{\bf #1} (#2) #3}
 \def\PRS #1 #2 #3 {Proc.~Roy.~Soc.~{\bf #1} (#2) #3}
 \def\IM #1 #2 #3 {Inv.~Math.~{\bf #1} (#2) #3}
 \def\JFA #1 #2 #3 {J.~Funkt.~Anal.~{\bf #1} (#2) #3}
 \def\LMP #1 #2 #3 {Lett.~Math.~Phys.~{\bf #1} (#2) #3}
 \def\IJMP #1 #2 #3 {Int.~J.~Mod.~Phys.~{\bf #1} (#2) #3}
 \def\FAA #1 #2 #3 {Funct.~Anal.~Appl.~{\bf #1} (#2) #3}
 \def\AP #1 #2 #3 {Ann.~Phys.~{\bf #1} (#2) #3}
 \def\MPL #1 #2 #3 {Mod.~Phys.~Lett.~{\bf #1} (#2) #3}

\def\ap{a_+}
\def\am{a_-}
\def\au{a_1}
\def\ad{a_2}
\def\d{\partial}

\def\du{\partial_u}
\def\dv{\partial_v}
\def\f{\phi}
\def\vf{\varphi}
\def\r{\rho}

\def\s{\sigma}

\def\l{\lambda}
\def\t{\tau}

\def\rg{\sqrt{-g}}

\def\eom{equation of motion\ }
\def\eoms{equations of motion\ }
\def\m{\mu}
\def\n{\nu}

\def\e{\epsilon}
\def\es{\epsilon(\s-\s')}
\def\ds{\delta(\s-\s')}
\def\dsp{\delta'(\s-\s')}
\def\dspp{\delta''(\s-\s')}
\def\dsppp{\delta'''(\s-\s')}
\def\a{\alpha}
\def\b{\beta}
\def\g{\gamma}
\def\gd{\gamma^2}
\def\gmd{\gamma^{-2}}
\def\rd{{\rm d}}
\def\th{{\rm th}}
\def\sh{{\rm sh}}
\def\ch{{\rm ch}}
\def\G{{\cal G}}
\def\la{\vert \l_\a \rangle}
\def\lb{\langle \l_\b \vert}
\def\vac{\vert 0\rangle}
\def\FL{{\cal F}_L}
\def\FR{{\cal F}_R}
\def\st{^{\ \star}}
\def\pr{\Pi_r}
\def\pt{\Pi_t}
\def\pf{\Pi_\f}
\def\Tb{{\bar T}}
\def\Vb{{\bar V}}

{ \chapter{ Introduction}}

The study of strings propagating on a curved background is a
difficult subject. In general our knowledge is restricted to
perturbation in the inverse string tension $\a'$
\REF\CP{C.G. Callan, D. Friedan, E.J. Martinec and M.J. Perry,
\NP B262 1985 593 .}
[\CP]. Sometimes the $\s$-model describing the string can be
shown to be equivalent to a known conformal field theory
\REF\WIT{E. Witten, \PR D44 1991 314 .}
[\WIT]. However, we do not
know the general classical solution to the equations of motion of a
string on a Schwarzschild background.
Even in conformal gauge, one has to solve the coupled non-linear
partial differential equations
$$
\du\dv X^\m +\Gamma^\m_{\n\r}(X) \du X^\n\dv X^\r =0
\eqn\uii$$
where $ \Gamma^\m_{\n\r}$ is the Christoffel symbol associated with
the background metric, and $u=\t+\s,\ v=\t-\s$ are light-cone
coordinates on the world-sheet. Of course, {\it particular} solutions
are easy to find: we can look e.g. for $\s$-independent solutions (no
oscillator excitations) and find that the string center of mass
describes the geodesics, well known for the Schwarzschild metric.
Starting from this special solution we can generate others,
exploiting the conformal invariance and making a conformal
transformation of the world-sheet coordinates, thereby obtaining
solutions depending on two arbitrary functions of one variable.

However, if we want to (canonically) quantize the theory,  and
eventually compute the partition function and the entropy of the string
on the black hole background we need the complete {\it general}
solutions. Obtaining the general solution of eq. \uii\ for the
Schwarzschild metric in 4D seems far beyond our present abilities, and one
seems forced to consider instead the two-dimensional (euclidean)
 black hole with metric
$$ {\rm d}s^2={\rm d}r^2+\th^2 r\, {\rm d}t^2\ .
\eqn\ui$$
Classical solutions can be obtained by exploiting the (classical)
equivalence with the gauged $Sl(2,{\bf R})/U(1)$ WZW-model [\WIT]. On the
other hand, quantization modifies the metric \ui\ by higher order
corrections in the WZW coupling constant ${1\over k}$. It might be useful
to have another model to investigate classical solutions and quantization.
It was noted some time ago
\REF\GS{J.-L. Gervais and M.V. Saveliev, \PL B286 1992 271 .}
[\GS] that the  equations of motion still are exactly solvable if one
adds another (``internal") string field with a particular
Liouville-type exponential interaction. The complete action of this
theory is
$$ S= {2\over \gd}\int \rd\s \rd\t
\left( \du r\dv r+\th^2 r\, \du t\dv t +\du \f\dv\f + \ch 2r\, e^{2\f}
\right) \ .
\eqn\uiii$$
The first two, $\f$-independent terms, are precisely the 2D black
hole sigma-model action, while the last two, $\f$-dependent terms
correspond to an ``internal" field $\f$ (or a flat third dimension) and
a tachyon potential $\ch 2r\, e^{2\f}$. The constant $\gd$ plays the role
of the Planck constant and will be seen later on to control the central
charge of the conformal algebra. As discussed below, this model is obtained
by gauging a {\it nilpotent} subalgebra of the Lie algebra $B_2$.

One may view \uiii\ as the conformal gauge version of the general
sigma-model
$$ S={1\over \gd} \int \rd^2 z \rg
\left[ {1\over 2} g^{\a\b} G_{\m\n}(X)\d_\a X^\m \d_\b X^\n
-T(X) +\Phi(X) R^{(2)} \right]
\eqn\uiv$$
where $g_{\a\b}$ is the world-sheet metric with curvature $R^{(2)}$,
$\Phi$ the dilaton, $G_{\m\n}$ the metric describing the space-time
background and $T$ the so-called tachyon potential. We have
$$ G_{\m\n}=\pmatrix{1&0&0\cr 0&\th^2r&0\cr 0&0&1\cr} \quad ,
\qquad T=-\ch 2r e^{2\f}\ .
\eqn\uv$$
In the conformal gauge action \uiii\  the dilaton $\Phi$ has
disappeared, but it should be remembered that the full (improved) stress
tensor obtained from \uiv\ by varying the metric $g_{\a\b}$ equals not only
the canonical (Noether) stress tensor $T_{\a\b}$ obtained from \uiii\
but also includes a contribution from the dilaton
$\sim(\eta_{\a\b}\d^2-\d_\a \d_\b)\Phi$.

In the present paper, I will be only concerned with the theory
defined by the action \uiii. It will turn out to be quite  interesting as a
conformal field theory in its own right. It is known as the non-abelian Toda
theory \REF\LS{A.N. Leznov and M.V. Saveliev, \CMP 89 1983 59 .} [\LS]
associated with the Lie algebra $B_2$ [\GS].  The general solution of the
equations of motion is in principle contained in ref. \GS\ where it is shown
how the solutions for an equivalent system of equations can be obtained from
the general scheme of ref. \LS. However, it is non-trivial to actually spell
out the solution and put it in a compact and useful form. This will be done
in section 2, after a brief review of the main results of refs. \GS, \LS. The
explicit solutions I obtain for the fields exhibit an amazing
factorization into relatively simple factors which makes it possible
to compute the three left-moving conserved quantities $T\equiv T_{++},
V^+\equiv V^+_{++}$ and $V^-\equiv V^-_{++}$ and express them in terms of
three functions $f_i(u)$ of one light-cone variable only. The same is true
for the three right-moving conserved quantities
$\Tb\equiv T_{--},
\Vb^+\equiv V^+_{--}$ and $\Vb^-\equiv V^-_{--}$ which are expressed in
terms of three functions $g_i(v)$ only.
 Although a priori
expected, the way this actually works is highly non-trivial and constitutes
a severe consistency check on the solution. Part of the algebraically
somewhat involved computations of this section is transferred into the
appendix B.

The complete set of conserved charges must form a closed
(Poisson bracket) algebra, since otherwise one would generate new
conserved quantities. Since the present theory has three fields
$r,t,\f$  one expects that the three (left-moving) conserved quantities $T,
V^+$ and $V^-$ form a complete (left-moving) set and hence a closed Poisson
bracket algebra. In section 3, I compute this Poisson bracket algebra using
the canonical Poisson brackets of the original fields $r,t,\f$. The $T$ and
$\Tb$ turn out to generate each a Virasoro algebra with classical central
charge $c\sim {1\over \gd}$. The Poisson brackets of $T$ with $V^+$ and $V^-$
show that the latter have conformal dimension 2 (no anomaly) while the
Poisson bracket of $V^\pm$ with $V^\pm$ or $V^\mp$ gives a non-linear,
non-local but closed expression of the three generators. The same applies to
the $--$ components $\Vb^\pm$.

Probably the most important step in solving an integrable model is to
obtain the Poisson brackets for the free fields of one variable only,
the $f_i(u)$ and the $g_i(v)$, since these or some suitable functions
therof will serve as a basis for quantization. Given the rather involved
transformation from the original fields $r,t,\f$ (and their momenta $\Pi_r,
\Pi_t, \Pi_\f$) to the $f_i(u), g_i(v)$ obtained in section 2  it would be a
formidable task to work out these Poisson brackets directly. There is,
however, an alternative simpler route. The conserved charges
$T, V^+, V^-$, resp. $\Tb, \Vb^+, \Vb^-$, are relatively simple
functionals of the $f_i(u)$, resp $g_i(v)$, only. It is not too
difficult to deduce the Poisson brackets of the $f_i$ and of the $g_i$
from those of the $T$'s and $V$'s. As
expected from experience with integrable models, the $f_i$ and $g_i$ Poisson
brackets are simple, but they can be made even simpler: by a further
(chiral) transformation they are turned into harmonic oscillator
Poisson brackets, $i\{ \vf^j_n,\vf^k_m\}\sim n \delta_{n,-m}
\delta^{j,k}$. This completes the classical solution of the theory
defined by \uiii, and will be the starting point for quantization
which I intend to discuss in a separate publication. Here, in section 4, I
only make a few remarks about quantization and  a couple of other
issues, like zero-modes and periodicity in
$\s$ of the solution, as well as how to implement periodicity of the
Euclidean (target space) time $t$, or the possible structure of the quantum
algebra of the conserved quantities. Another interesting topic is the
associated hierarchy of integrable partial differential equations leading
probably to some non-abelian version of the KP hierarchy. However, none of
these issues will be fully solved here.

{\chapter{The classical solution}}

\section{Review of known facts}

The \eoms obtained from the action \uiii\ read
$$\eqalign{
\du\dv r&={\sh r\over \ch^3r}\du t\dv t +\sh 2r\, e^{2\f}\cr
\du\dv t&=-{1\over \sh r\, \ch r}\left( \du r\dv t+\du t\dv r\right)\cr
\du\dv\f&=\ch 2r\, e^{2\f} \ .}
\eqn\di$$
Using these equations of motion it is completely straightforward to show
that the following three quantities are conserved [\GS]:
$$\eqalign{
T\equiv T_{++}&=(\du r)^2+\th^2r\, (\du t)^2 +(\du \f)^2-\du^2\f\cr
V^\pm\equiv
V^\pm_{++}&={1\over \sqrt{2}} \left( 2\du\f-\du\right)
\left[e^{\pm i\n}\left(\du r\pm i \th r\, \du t\right)\right]\cr }
\eqn\dii$$
i.e.
$$\dv T=\dv V^\pm=0\ .
\eqn\diii$$
Here $\n$ is defined by
$$\dv\n=\ch^{-2}r\, \dv t\quad , \qquad \du\n=(1+\th^2r)\du t
\eqn\div$$
where the integrability condition is fulfilled due to the \eoms \di.
Similarly one has for the $--$ components $\du \Tb=\du \Vb^\pm=0$
where
$$\eqalign{
\Tb\equiv T_{--}&=(\dv r)^2+\th^2r\, (\dv t)^2 +(\dv \f)^2-\dv^2\f\cr
\Vb^\pm\equiv V^\pm_{--}&={1\over \sqrt{2}} \left( 2\dv\f-\dv\right)
\left[e^{\mp i\m}\left(\dv r\mp i \th r\, \dv t\right)\right]\cr }
\eqn\dvv$$
and\foot{
The $\Vb^\pm $ are not given in ref. \GS\ but can easily be guessed, or
actually deduced from $V^\pm$ using the star operation defined below.}
$$\du\m=\ch^{-2}r\, \du t\quad , \qquad \dv \m=(1+\th^2r)\dv t\ .
\eqn\dvi$$

The essential point for solving the \eoms \di\ is to realize that they
can be derived from a Lax pair. Following Gervais and Saveliev
[\GS]\foot{There are some obvious misprints in ref. \GS\ which are
corrected here.}, one first introduces fields $\au, \ad, \ap$ and $\am$
subject to the following \eoms
$$
\eqalign{
&\du\dv \au=-2(1+2\ap\am)e^{-\au}\cr
&\du\dv(2\ad-\au)+2\du(\dv\ap \am)=0\cr
&\du\left(e^{\au-2\ad}\dv\ap\right)=2\ap e^{-2\ad}\cr
&\du\left[e^{-\au+2\ad}\left(\dv\am-\am^2\dv\ap\right)\right]=2 \am
(1+\ap\am) e^{-2\au+2\ad}\ .\cr }
\eqn\dvii$$
It is then straightforward, although a bit lengthy, to show that we can
identify
$$\eqalign{
\f&=-{1\over 2}\au\cr
\sh^2 r&=\ap\am\cr
\du t&={1\over 2i}\left[(1+2\ap\am){\du\am\over \am}- {\du\ap\over \ap}
-2(1+\ap\am)\du(\au-2\ad)\right]\cr
\dv t&={i\over 2}\left[(1+2\ap\am){\dv\ap\over \ap}- {\dv\am\over \am}
\right]\cr}
\eqn\dviii$$
i.e. with these definitions  the equations \dvii\ and \di\ are equivalent.
Again, the integrability condition for solving for $t$ is given by the
equations \dvii.

The advantage of equations \dvii\ over equations \di\ is that they follow,
as will be shown next, from the Lie algebraic formulation
$$\du( g_0^{-1}\dv g_0)=[J_-\, ,\, g_0^{-1}J_+ g_0]\ .
\eqn\dix$$
Note, that this is obviously equivalent to the Lax
representation
$$\eqalign{
&[ \du-{\cal A}_u\, ,\, \dv-{\cal A}_v ]=0\ ,\cr
&{\cal A}_u =-g_0^{-1} J_+ g_0 \cr
&{\cal A}_v=\dv g_0^{-1} g_0 -J_- \ .\cr }
\eqn\lax$$
Here the relevant algebra is $B_2$ with
generators $h_1, h_2$ (Cartan subalgebra) and $E_{e_1}, E_{e_2}, E_{e_1-e_2},
E_{e_1+e_2}$ and their conjugates $E_\a^+=E_{-\a}$; see appendix A for their
definitions in terms of fermionic oscillators. Then $H=2h_1+h_2,
J_+=E_{e_1}$ and $J_-=E_{-e_1}$ span an $A_1$ subalgebra. $H$ induces a
gradation on $B_2$. The gradation 0 part ${\cal G}_0$  is spanned by $h_1,
h_2, E_{e_2}$ and $E_{-e_2}=E_{e_2}^+$. The corresponding group elements
$g_0\in G_0$ can be parametrized as
$$g_0=\exp(\ap E_{e_2})\exp(\am E_{e_2}^+)\exp(\au h_1 +\ad h_2)\ .
\eqn\dx$$
Using the commutation relations of $B_2$, it is then easy to show  that with
these definitions of $g_0$ and $J_\pm$, equation \dix\ is equivalent to
equations \dvii.

Now, following ref. \GS, I will show how to obtain the solution of
equation \dix. What follows applies to any Lie algebra $\G$, not only $B_2$.
Suppose $\G$ has a grading under which it decomposes as $\G=\G_+\oplus \G_0
\oplus \G_-$. Then, every element of the corresponding group $G$ has a
unique Gauss decomposition
$$ g=g_- g_0 g_+\ .
\eqn\dxi$$
(Of course, there is also a similar decomposition $\tilde g_+ \tilde g_0
\tilde g_-$.) Take some fixed elements $J_\pm \in \G_\pm$ and let for
every $g_0\in G_0$ the elements $g_\pm\in G_\pm$ be solutions of the
differential equations
$$\eqalign{
\du g_+^{-1}&=g_+^{-1} (g_0^{-1} J_+ g_0)\cr
\dv g_-&=g_- (g_0 J_- g_0^{-1})
\ .\cr}
\eqn\dxii$$
Let $g=g_- g_0 g_+$. Equations \dxii\ define $g_+$ and $g_-$ up to  right
(resp. left) multiplications by a group element depending only on $v$ (resp.
$u$). Thus $g$ is defined up to $g\to F(u) g H(v)$. It follows from these
differential equations that
$$\eqalign{
\du (g^{-1}\dv g)&=g_+^{-1}\left\{    \du(g_0^{-1}\dv g_0)
 -[J_-\, ,\, g_0^{-1} J_+ g_0]\right\} g_+\cr
\dv (\du g g^{-1})&=-g_- g_0 \left\{    \du(g_0^{-1}\dv g_0)
 -[J_-\, ,\, g_0^{-1} J_+ g_0]\right\} g_0^{-1} g_-^{-1}
\cr}
\eqn\dxiii$$
and
$$\du (g^{-1}\dv g)=0 \quad {\rm and} \quad  \dv (\du g g^{-1})=0
\eqn\dxiv$$
 are both
equivalent to equation \dix. The general solution to  equations \dxiv\ is
$g=g_L(u) g_R(v)$. But each group element $g_L$ and $g_R$ has again a Gauss
decomposition
$$\eqalign{
g_L(u)&=g_{L-}(u)g_{L0}(u)g_{L+}(u)\ ,\cr
g_R(v)&=g_{R-}(v)g_{R0}(v)g_{R+}(v)\cr }
\eqn\dxv$$
so that
$$g=g_{L-}(u)g_{L0}(u)g_{L+}(u)g_{R-}(v)g_{R0}(v)g_{R+}(v)\ .
\eqn\dxvi$$
On the other hand we also have the decomposition \dxi\ and $g_+$ and $g_-$
must obey the differential
equations \dxii. The latter translate into
$$\eqalign{
\du g_{L+}(u)=-{\cal F}_L(u) g_{L+}(u) \quad &,
\quad {\cal F}_L(u) =g_{L0}^{-1}(u) J_+ g_{L0}(u)\ ,\cr
\dv g_{R-}(v)= g_{R-}(v){\cal F}_R(v) \quad &,
\quad {\cal F}_R(v) =g_{R0}(v) J_- g_{R0}^{-1}(v)\ .\cr
}
\eqn\dxx$$
The strategy then is
\pointbegin
Pick some arbitrary $g_{L0}(u),\ g_{R0}(v)\in G_0$.
\point
Compute the solutions $g_{L+}(u)$ and $g_{R-}(v)$ from the first order
ordinary differential equations \dxx.
\point
Let
$$
\Gamma=g_{L0}(u) g_{L+}(u) g_{R-}(v) g_{R0}(v)
\eqn\dxxi$$
and choose a basis $\la$ of states annihilated by $\G_+$. Then using \dxi\
and \dxvi\ we have
$$G_{\a\b}\equiv \lb g_0 \la=\lb \Gamma \la\ .
\eqn\dxxii$$
This yiels all matrix elements of $g_0$, solution of equation \dix, which
in turn, as shown above, yields the solution for the $\au,\ad,\ap$ and
$\am$.
\par

\section{The explicit solution}

%CHANGE $g_{1,2}\to -g_{1,2}, g^+\to -g_-, g^-\to -g_+$
%w.r.t. notation in my computations.

I now apply the general strategy to the case of present interest with Lie
algebra $B_2$. To carry out point 1, parametrize
$$\eqalign{
g_{L0}&=\exp\left( -f_1(u) h_1-f_2(u) h_2 \right) \
\exp\left( -f_-(u) E_{-e_2} \right)\
\exp\left( -f_+(u) E_{e_2} \right) \cr
g_{R0}&=\exp\left( -g_+(v) E_{-e_2} \right)\
\exp\left( -g_-(v) E_{e_2} \right)\
\exp\left( -g_1(v) h_1-g_2(v) h_2 \right)
 \cr}
\eqn\dxxx$$
so that
$$g_{R0}=g_{L0}^+ \Big\vert_{f\to g}
\eqn\dxxxi$$
(where  the $f_i, g_i$ are treated as real under hermitian conjugation).
More generally, I define a star operation, denoted by $\st$, which
essentially interchanges left and right movers. More precisely, it
includes hermitian conjugation (treating $f_i$ and $g_i$ as real), the
interchange of $f_i$ and $g_i$, and replaces $\int{\rm d}u$ by $-\int {\rm
d}v$ as well as  $\du$ by $-\dv$ and vice versa:
$$X\st\equiv X^+\Big\vert_{f_i(u)\leftrightarrow g_i(v),\, \int {\rm d}u
\leftrightarrow -\int {\rm d}v,\, \du \leftrightarrow -\dv}\ .
\eqn\dxxxia$$
Note that
$$(X\st)\st = X\ .
\eqn\dxxxib$$
Then $\FL$ and $\FR$ are easily obtained from their definition \dxx\ with
$J_\pm=E_{\pm e_1}$:
$$\eqalign{
\FL(u)&=e^{f_1}\left[ (1+2f_+f_-)E_{e_1}-2f_-E_{e_1-e_2}+2f_+(1+f_+f_-)
E_{e_1+e_2}\right]\cr
\FR(v)&=\FL^+\Big\vert_{f(u)\to g(v)}=\FL\st\ .\cr}
\eqn\dxxxii$$
Note that they do not depend on the functions $f_2$ and $g_2$.  The
group elements $g_{L+}(u)$ and $g_{R-}(v)$ are given by the solutions of
\dxx\ as %
$$\eqalign{
g_{L+}(u)&=1-\int^u {\rm d}u_1\, \FL(u_1)
+\int^u {\rm d}u_1 \int^{u_1} {\rm d}u_2\, \FL(u_1)\FL(u_2)-\ldots \cr
g_{R-}(v)&=1+\int^v {\rm d}v_1\, \FR(v_1)
+\int^v {\rm d}v_1 \int^{v_1} {\rm d}v_2\, \FR(v_2)\FR(v_1)+\ldots \cr
}
\eqn\dxxxiii$$
so that
$$g_{R-}(v)=g_{L+}^+(u)\Big\vert_{f\to g,\, \int\rd u\to -\int\rd
v}=g_{L+}\st \quad \Rightarrow \quad g_{R-}g_{R0}=(g_{L0}g_{L+})\st
%(g_{L0}g_{L+})^+\Big\vert_{f\to g, \int\to -\int}
\ .
\eqn\dxxxiv$$
This completes step 2.

All one has to do to implement step 3 is to compute
$\vert \psi_\a(v)\rangle =g_{R-}g_{R0}\la$ for the various $\la$ annihilated
by $\G_+$, i.e. annihilated by $E_{e_1},\  E_{e_1 +e_2}$ and $E_{e_1-e_2}$.
Indeed, once the $\vert \psi_\a(v)\rangle $ are known, one has
$$G_{\b\a}=\langle
\chi_\b(u)\vert \psi_\a(v)\rangle
\eqn\dxxxiva$$
where
$\langle \chi_\a(u)\vert=\vert \psi_\a(v)\rangle\st$. As in ref. \GS, I
choose
$$\vert \l_1\rangle = b_1^+\vac\quad , \quad
\vert \l_2\rangle = b_2^+ b_1^+\vac\quad , \quad
\vert \l_3\rangle = b_0^+ b_1^+\vac\ .
\eqn\dxxxv$$
The actual computation of the $\vert\psi_\a\rangle$ is rather cumbersome,
although straightforward. To write the results in a more compact way,
introduce the functions of one variable
$$\eqalign{
F_1(u)&=-\int^u e^{f_1} (1+2f_+f_-)\cr
F_2(u)&=2\int^u e^{f_1} f_-\cr
F_3(u)&=-2\int^u e^{f_1}f_+(1+f_+f_-) \cr}
\eqn\dxxxvi$$
as well as
$$F_+=F_1+f_+ F_2\quad , \quad F_-=F_3-f_+ F_1
\eqn\dxxxvii$$
and
$$G_i(v)=F_i(u)\st\quad , \quad
G_\pm(v)=F_\pm(u)\st\ .
\eqn\dxxxviii$$
The three vectors $\vert \psi_\a\rangle$ then are
$$\eqalign{
\vert\psi_1\rangle &=
g_{R-}g_{R0} b_1^+\vac \cr
&= e^{-g_1}\Big\{  b_1^++\sqrt{2} G_1 b_0^+ +G_2
b_2^+ +G_3b_{-2}^+
-(G_1^2+G_2G_3)b_{-1}^+\Big\}\ \vac \ ,}
\eqn\dxxxix$$
$$\eqalign{\vert\psi_2\rangle =
g_{R-}g_{R0} &b_2^+ b_1^+\vac \cr
= e^{-2g_2}\Big\{
& b_2^+ b_1^+ -\sqrt{2}g_+b_0^+b_1^+-g_+^2b_{-2}^+b_1^+\cr
&+\sqrt{2}G_+b_2^+b_0^+-\sqrt{2}g_+G_-b_0^+b_{-2}^+\cr
&+(G_--g_+G_+)b_1^+b_{-1}^++(G_-+g_+G_+)b_2^+b_{-2}^+\cr
&+G_+^2b_{-1}^+b_2^+-G_-^2b_{-1}^+b_{-2}^++\sqrt{2}G_+G_-b_0^+b_{-1}^+
\Big\}\ \vac \ , \cr
}
\eqn\dxxxx$$
and
$$\eqalign{\vert\psi_3\rangle =
g_{R-}&g_{R0} b_0^+ b_1^+\vac \cr
= e^{-g_1}\Big\{
& -\sqrt{2}g_-b_2^+b_1^++(1+2g_+g_-)b_0^+b_1^+
+\sqrt{2}g_+(1+g_+g_-)b_{-2}^+ b_1^+\cr
&-(2g_-G_++G_2)b_2^+b_0^+ +[(1+2g_+g_-)G_--g_+G_1]b_0^+b_{-2}^+\cr
&+\sqrt{2}[(1+g_+g_-)G_+-g_-G_-]b_1^+b_{-1}^+\cr
&-\sqrt{2}[g_-G_-+g_+g_-G_++g_+G_2]b_2^+b_{-2}^+\cr
&-\sqrt{2}(g_-G_+^2+G_+G_2)b_{-1}^+b_2^+
+\sqrt{2}(g_-G_-^2-G_-G_1)b_{-1}^+b_{-2}^+\cr
&-(2g_-G_+G_-+G_-G_2-G_+G_1)b_0^+b_{-1}^+
\Big\}\ \vac \ .\cr
}
\eqn\dxxxxi$$
{}From these expressions one obtains the matrix elements $G_{\a\b}=\langle
\chi_\a(v)\vert\psi_\b(u)\rangle$. Obviously,
$G_{12}=G_{13}=G_{21}=G_{31}=0$. The matrix element $G_{11}$ is relatively
simple, but $G_{22}, G_{33}$ and $G_{23}, G_{32}$ look discouragingly
complicated at first sight (e.g. $G_{33}$ contains about 150 terms).
However, one realizes that the $G_{\a\b}$ factorize into products of
simpler quantities:
$$\eqalign{
G_{11}&=e^{-f_1-g_1} Z\cr
G_{22}&=e^{-2f_2-2g_2}X^2\cr
G_{23}&=\sqrt{2}e^{-2f_2-g_1}XV\cr
G_{32}&=\sqrt{2}e^{-f_1-2g_2}XW\cr
G_{33}&=-G_{11}+2e^{-f_1-g_1}XY\ .\cr
}
\eqn\dxxxxiii$$
The quantities $X,Y,Z$ and $V,W$ are
$$\eqalign{
X&=1+f_+g_++F_+G_++F_-G_-\cr
Y&=(1+f_+f_-)(1+g_+g_-)+f_-g_-+(F_1-f_-F_-)(G_1-g_-G_-)\cr
&\phantom{=}+(F_2+f_-F_+)(G_2+g_-G_-)\cr
Z&=1+2F_1G_1  +F_2G_2+F_3G_3+(F_1^2+F_2F_3)(G_1^2+G_2G_3)\cr
V&=-g_--f_+-g_+g_-f_+-g_-F_+G_+-g_-F_-G_-+F_-G_1-F_+G_2\cr
W&=-f_--g_+-f_+f_-g_+-f_-F_+G_+-f_-F_-G_-+F_1G_--F_2G_+\ .\cr
}
\eqn\dxxxxiv$$
Note that under the star operation
$$X\st=X\ ,\  Y\st=Y\ ,\ Z\st=Z\quad , \quad V\st=W\ ,\ W\st=V\ .
\eqn\dxxxxv$$

At this point one has to recall that the $G_{\a\b}$ are the matrix
elements of $g_{L0}g_{L+}g_{R-}g_{R0}$ between $\langle\l_\a\vert$ and
$\vert\l_\b\rangle$. But according to the general discussion of the
previous subsection (see point 4 of the strategy) they equal
$\langle\l_\a\vert g_0\vert\l_\b\rangle$. If one uses the parametrization
\dx\ of $g_0$ in terms of $\au,\ad,\ap$ and $\am$ (each depending on $u$
and $v$, in contrast with the $f_i(u)$ and $g_i(v)$) one finds
$$\eqalign{
G_{11}&=e^{a_1} \cr
G_{22}&=e^{2a_2}(1+\ap\am)^2\cr
G_{23}&=\sqrt{2}e^{a_1}\ap(1+\ap\am)\cr
G_{32}&=\sqrt{2}e^{2a_2}\am(1+\ap\am)\cr
G_{33}&=-e^{a_1}+2e^{a_1}(1+\ap\am)\ .\cr
}
\eqn\dxxxxvi$$
Comparing equations \dxxxxvi\ and \dxxxxiii, one obtains the
$\au,\ad,\ap,\am$ in terms of the $V,W,X,Y,Z$. Note that we have 5
equations for only 4 $a$'s. This implies a relation between the $V,\ldots Z$
as will be shown below. Note also that equations \dxxxxvi\ have the same
factorized form as \dxxxxiii\ which allows for immediate identifications like
$e^{\au}=e^{-f_1-g_1}Z$ and $1+\ap\am={XY\over Z}$. The complete solution is
$$\eqalign{
e^{a_1}&= e^{-f_1-g_1}Z\cr
e^{-a_2}&=\pm e^{f_2+g_2}{Y\over Z}\cr
\ap&=e^{f_1-2f_2}{V\over Y}\cr
\am&=e^{2f_2-f_1}{YW\over Z}\cr
}
\eqn\dxxxxvii$$
while the fifth equation is the relation
$$XY-Z=VW
\eqn\dxxxxviii$$
which is easily verified. Some other relations satisfied by the $V,\ldots
Z$ and their derivatives are given in appendix B. It is also noteworthy
that under the star operation
$$\eqalign{
(e^{\au})\st=e^{\au}\quad &, \quad (e^{-\ad})\st=e^{-\ad}\cr
(\ap)\st=e^{2\ad-\au}\am\quad &, \quad (\am)\st=e^{\au-2\ad}\ap \cr}
\eqn\dxxxxix$$
so that $(\ap\am)\st=\ap\am$. It is then seen from \dviii\ that the fields
$\f, \, r$ and $t$ are invariant under the star operation. Actually, since
one only determines $\sh ^2r$ (or $\ch^2 r$) one has a sign ambiguity for
$r$ and one could decide that $r$ changes sign under the star operation.
(However from the euclidean black hole point of view, $r$ should be always
non-negative, hence is invariant.) Also $t$ is invariant if one assumes $i\st
=  -i$ (as would be normally implied by hermitian conjugation). If one
considers the Minkowskian continuation $\theta=i t$ instead, one sees that
the star operation corresponds to Minkowskian time reversal.

{}From equations \dviii\ one immediately has
$$\eqalign{
\f&={1\over 2}\left( f_1+g_1-\log Z\right)\cr
\sh^2 r&={VW\over Z}={XY\over Z}-1\cr
}
\eqn\dxxxxx$$
while some more work is needed to integrate the two equations for $t$. Using
equations (B.2) and (B.3) from appendix B one finds
$$\eqalign{
-2i\dv t&=2{W\dv V-X\dv Y\over Z}+\dv\log {VZ\over W}\cr
&=-2g_-g_+'+\dv\log {V\over W}\cr }
\eqn\dxxxxxi$$
and
$$
-2i\du t=2f_-f_+'+\du\log {V\over W}
\eqn\dxxxxxii$$
which can be integrated to give
$$
t=t_0+i\int^u f_-f_+'-i\int^v g_-g_+' +{i\over 2} \log {V\over W}
\eqn\dxxxxxiii$$
(where it is again obvious that $t\st=t$). This completes the explicit
solution of the field equations of motion.

Before going on, let me make a remark about the reality properties of the
solution.  For real functions $f_i,\ g_i$ the euclidean time $t-t_0$ is
purely imaginary (provided $V/W>0$\foot{
If $V/W<0$ the corresponding $(i/2)i\pi$ can be absorbed into $t_0$.}) so
that the Minkowski time $\theta=it$ is real. Thus real $f_i, g_i$ are
appropriate for a Minkowskian ``target space" time. On the other hand, if we
use euclidean ``world-sheet" coordinates $u,v$ so that $u^*=v$ (here $^*$
means complex conjugation), we can use complex $f_i, g_i$ such that
$f_i(u)^*=g_i(v)$, so that the exchange of $f$ and $g$ in the star operation
is part of the hermitian conjugation. Then $V\st=W$ simply becomes $V^*=W$
and $\log {V\over W}$ is purely imaginary. It follows that in this case
$t-t_0$ is real. So this is the appropriate setting for euclidian ``target
space" time $t$. In both cases, $r$ and $\f$ are real.

One might want to check directly that the equations of motion are indeed
satisfied. This can be done. For example, the
$\au$-equation of motion reduces to equation (B.4). Rather than
verifying the $t$-\eom directly, one may observe that the integrability of
the above equations \dxxxxxi\ and \dxxxxxii\ is a severe consistency check
on the solution. Moreover, in the next subsection, I express the conserved
quantities $T$ and $V^\pm$ in terms of the $f_i(u)$ only. This
would almost certainly fail if there were only the slightest error in the
solution \dxxxxiv, \dxxxxvii.

\section{The conserved quantities}

Using the \eoms for $\f,\, r$ and $t$ it was shown above that the
quantities $T\equiv T_{++}$ and $V^\pm\equiv V^\pm_{++}$ are conserved, i.e.
can only depend on $u$. This means that they must be expressible entirely in
terms of the $f_i(u)$'s. Given the complexity of the solutions \dxxxxx,
\dxxxxxiii\ and \dxxxxiv\ this is highly  non-trivial, and, as already
mentioned, constitutes a severe consistency check. The same considerations
apply to $\Tb\equiv T_{--}$ and $\Vb^\pm\equiv V^\pm_{--}$.

To begin with, I consider $T_{\pm\pm}$ as given by \dii\ and \dvv. When
expressed in terms of $\au, \ad, \ap$ and $\am$ they read
$$\eqalign{
\Tb\equiv T_{--}&={1\over 4}(\dv \au)^2+{1\over 2}\dv^2
\au+\dv\ap(\dv\am-\am^2\dv\ap)\cr
T\equiv T_{++}&={1\over 4}(\du \au)^2+{1\over 2}\du^2 \au\cr
&\phantom{=}+\du\left(
e^{2\ad-\au}\am\right)\left[ \du  \left( e^{\au-2\ad}\ap\right)-\left(
e^{\au-2\ad}\ap\right)^2 \du\left( e^{2\ad-\au}\am\right)\right]\cr
&=T_{--}\st\ .\cr
}
\eqn\ddi$$
It is easier to compute $T_{--}$ first. Using equation \dxxxxvii\ and
performing straightforward algebra one obtains
$$T_{--}={1\over 4}(g_1')^2-{1\over 2}g_1'' +{\dv^2 Z-g_1'\dv Z\over 2Z}
-\left( {\dv Z\over 2Z}\right)^2 +t_{--}
\eqn\ddii$$
with
$$\eqalign{
t_{--}={1\over Y^2Z^2}&[Y\dv V-V\dv Y]\times \cr
\times &[Y(Z\dv W-W\dv Z)-W^2 (Y\dv V-V\dv Y)+WZ\dv Y]\ .\cr   }
\eqn\ddiii$$
Using the relation \dxxxxviii\ and its $\dv$ derivative one can simplify
$t_{--}$ to
$$t_{--}={1\over Z^2}(Y\dv V-V\dv Y)
(X\dv W-W\dv X)\ .
\eqn\ddiv$$
Inserting relations (B.5) and (B.6) of appendix B ($\a$ and $\b$ are
defined in this appendix) yields
$$t_{--}=g_+'(g_-'-g_-^2g_+')-{g_+'\b+(g_-'-g_-^2g_+')\a\over Z}+{\a\b\over
Z^2}\ .
\eqn\ddv$$
Combining this formula with \ddii\  one finds for $T_{--}$
a piece ${1\over 4}(g_1')^2-{1\over 2}g_1''+  g_+'(g_-'-g_-^2g_+')$ that
manifestly only depends on $v$, plus
$$\Delta={\dv^2 Z-g_1'\dv Z\over 2Z}
-\left( {\dv Z\over 2Z}\right)^2 -{g_+'\b+(g_-'-g_-^2g_+')\a\over Z}
+{\a\b\over Z^2}\ .
\eqn\ddvi$$
But this vanishes by relation (B.8), so that finally
$$T_{--}={1\over 4}(g_1')^2-{1\over 2}g_1''+  g_+'(g_-'-g_-^2g_+')\ .
\eqn\ddvii$$
This is again a very simple expression, surprisingly similar to \ddi.
However, in \ddi\ the $a$'s are interacting fields depending on $u$ and $v$
in a very complicated way, while in \ddvii\ the $g_i(v)$ are functions of
$v$ only. The $++$ component of $T$ is obtained by the star operation:
$$T_{++}={1\over 4}(f_1')^2-{1\over 2}f_1''+  f_+'(f_-'-f_-^2f_+')\ .
\eqn\ddviii$$
In the next section, I will make a further (although much simpler) field
redefinition $f_i(u)\to \varphi_i(u)$ that brings $T_{++}$ into the form
$\sim (\d\varphi)^2+\d^2\varphi$, which is the standard form for a
conformal field theory energy momentum tensor.

The next task is to compute the $V^\pm$ and $\Vb^\pm$ given by \dii\ and
\dvv. First, one has to integrate \div\ and \dvi\ for $\nu$ and $\mu$. Using
relations (B.2) and (B.3) this can be done with the result
$$\eqalign{
\nu&=2i\int^u f_-f_+' + {i\over 2}\log{VX\over WY}\cr
\mu&=-2i\int^v g_-g_+' - {i\over 2}\log{WX\over VY}\ .\cr
}
\eqn\ddix$$
Inserting this and the solutions for $\f, r$ and $t$ one obtains after
some simple algebra, e.g. for $V^+$:
$$V^+={e^{f_1}\over 2\sqrt{2} Z} \du\, \left[
\exp\left( -f_1-2\int^u f_-f_+'\right) {1\over X}\,
(2Z\du W-W\du Z-2WZf_-f_+')\right]\ .
\eqn\ddx$$
Then using (B.3) and \dxxxxviii\ this can be written as
$$V^+={e^{f_1}\over \sqrt{2} Z} \du\, \left[
\exp\left( -f_1-2\int^u f_-f_+'\right)  (Y\du W-W\du Y) \right]\ .
\eqn\ddxi$$
Making further use of the relation obtained from (B.6) by the star
operation and equation (B.10) one finally arrives at
$$V^+={1\over \sqrt{2} } (f_1'-\du) \, \left[
\exp\left( -2\int^u f_-f_+'\right)  (f_-'-f_-^2f_+') \right]\ .
\eqn\ddxii$$
Again, one should compare this with the starting point \dii, but now
\ddxii\ is {\it manifestly} only a function of $u$. $V^-$ is obtained
along similar lines and I only give the result:
$$V^-={1\over \sqrt{2} } (f_1'-\du) \, \left[
\exp\left( +2\int^u f_-f_+'\right)  f_+' \right]\ .
\eqn\ddxiii$$
The $\Vb^{\pm}$ are obtained from the $V^\pm$ by the star
operation:
$$\eqalign{
\Vb^+&={1\over \sqrt{2} } (g_1'-\dv) \, \left[
\exp\left( -2\int^v g_-g_+'\right)  (g_-'-g_-^2g_+') \right]\cr
\Vb^-&={1\over \sqrt{2} } (g_1'-\dv) \, \left[
\exp\left( +2\int^v g_-g_+'\right)  g_+' \right]\ .\cr
}
\eqn\ddxiv$$
In conclusion, through the subtle ``magic" of integrable models, the
conserved quantities manage to be simple expressions of the functions
$f_i(u)$, resp. $g_i(v)$ of one variable only. The $V^{\pm}, \Vb^\pm$ depend
in a non-local way on the $f_i, g_i$. In the next section, motivated by the
Poisson brackets of the $f_i$'s ($g_i$'s), I make a non-local
transformation $f_i(u)\to \varphi_i(u)$ (and similarly for the $g$'s). The
$V^\pm$ then depend locally on the $\varphi_i(u)$.

Many more questions remain to be investigated at the level of the classical
solution to the equations of motion, and I will mention some of them in
section 4. Now, however, I turn to the discussion of the Poisson brackets
derived from the canonical structure.

\chapter{Symplectic structure and constraint algebra}

In this section I will determine the Poisson brackets of the various fields
and conserved quantities (constraints)\foot{
Since $T_{++}$ and $V^\pm_{++}$ are conserved under evolution of the
light-cone ``time" $v$ , they can be considered as constraints imposed on
the initial data. If one considered a Hamiltonian formalism of the theory
one would discover that $T_{\pm\pm}$ indeed appear as constraints conjugate
to Lagrange multipliers that play the role of world-sheet shift and lapse
functions. This leaves open the interpretation of the $V^\pm$. The answer
is probably closely related to the problem of $W$-gravity, and I expect
that the $V^\pm$ are constraints related to symmetries of the target space.
In any case I will refer to the $T$ and $V^\pm$ as constraints.}
encountered in the previous section. The program is
the following.

First, in subsection 3.1, using the canonical Poisson brackets of the fields
$r,\, t$ and $\f$ and their momenta, I compute the Poisson bracket algebra
of the conserved quantities $T$ and $V^\pm$ (constraints) when expressed in
terms of  $r,\, t$ and $\f$ and their derivatives. I will find that $T$
behaves as a stress tensor in conformal field theory and obeys the usual
Poisson bracket version of the Virasoro algebra with classical central
charge. The Poisson bracket of $T$ with $V^\pm$ just shows that $V^\pm$ are
conformally primary fields of weight 2. The Poisson bracket of $V^\pm$ with
$V^\pm$ or $V^\mp$ is more interesting (and more difficult to obtain). On
dimensional grounds one expects that the bracket of $V^\pm$ with $V^\mp$ can
contain $T$ and a central term, but also a term $V^\pm V^\mp$. All of them
do indeed appear. Due to the $U(1)$ charges $\pm 1$ and $0$ we can assign to
$V^\pm$ and $T$, the Poisson bracket of $V^\pm$ with $V^\pm$ can only give a
$V^\pm V^\pm$ term.

Then, in principle one could deduce the Poisson brackets
of the $f_i$ and $g_i$ through the transformation induced by the classical
solution \dxxxxx, \dxxxxxiii.
More precisely, one would have to allow formally that the $f_i$ and $g_i$
depend both on $u$ and $v$ since one has to consider the full phase space
and not only the manifold of solutions to the equations of motion.
Nevertheless, equations \dxxxxx\ and \dxxxxxiii, as well as their time
derivatives, constitute a phase space transformation from $r(\t,\s),\,
t(\t,\s),\, \f(\t,\s)$ and their momenta $\pr(\t,\s),\, \pt(\t,\s),\,
\pf(\t,\s)$ to new phase space variables $f_i(\t,\s),\ g_i(\t,\s)$. (Of
course, the equations of motion still imply $\dv f_i=\du g_i=0$.) In
practice, this would be very complicated to implement. It is much simpler
to use the Poisson brackets of the $T$ and $V^\pm$ derived before, and then
consider the $T$ and $V^\pm$ (or $\Tb$ and $\Vb^\pm$) as given in terms of
the $f_i$ only (or $g_i$ only).  Thus one does the phase space
transformation in two steps: $r(\t,\s),\,
t(\t,\s),\, \f(\t,\s),\, \pr(\t,\s),\, \pt(\t,\s),\,
\pf(\t,\s)\ \to\ T_{\pm\pm}(\t,\s),\,    V^+_{\pm\pm}(\t,\s),\,
V^-_{\pm\pm}(\t,\s)\    \to\ f_i(\t,\s),\, g_i(\t,\s)$
This yields the Poisson brackets of the $f_i$ and of the $g_i$ in a
relatively easy way (subsection 3.2). These Poisson bracket are simple, and
a final (non-local) transformation (subsection 3.3) turns them into standard
harmonic oscillator Poisson brackets.

\section{The constraint algebra}

Recall that the theory under consideration is based on the action \uiii.
Writing $u=\t+\s,\ v=\t-\s$, it becomes (as usual, a dot denotes $\d_\t$
while a prime denotes $\d_\s$)
$$ S= {1\over \gd}\int \rd\t \rd\s
\left[ {1\over 2} (\dot r^2-r'^2)+{1\over 2}\th^2 r\, (\dot t^2-t'^2)
+{1\over 2} (\dot \f^2-\f'^2) + 2\, \ch 2r\, e^{2\f} \right] \ .
\eqn\ti$$
The constant $\gd$ can be viewed as the Planck constant $2\pi \hbar$,
already included ino the classical action, or merely as a coupling
constant. The canonical momenta than are
$$\pr=\gmd\, \dot r\quad , \quad \pt=\gmd \th^2 r\, \dot t\quad , \quad
\pf=\gmd\, \dot\f
\eqn\tii$$
and the canonical (equal $\t$) Poisson brackets are
$$\eqalign{
\{r(\t,\s)\, ,\, \pr(\t,\s')\} &=
\{t(\t,\s)\, ,\, \pt(\t,\s')\} =
\{\f(\t,\s)\, ,\, \pf(\t,\s')\} = \ds\cr
\{r(\t,\s)\, ,\, r(\t,\s')\} &=
\{r(\t,\s)\, ,\, t(\t,\s')\} = \ldots = 0\cr
\{\Pi_i(\t,\s)\, ,\, \Pi_j(\t,\s')\}&=0\ . \cr
}
\eqn\tiii$$
It follows that
the only non-zero equal $\t$ Poisson brackets are
$$\eqalign{
\{r(\t,\s)\, ,\, \dot r(\t,\s')\} =\gd \ds\quad &, \quad
\{t(\t,\s)\, ,\, \dot t(\t,\s')\} ={\gd \over \th^2 r}\ds\ ,\cr
\{\f(\t,\s)\, ,\, \dot \f(\t,\s')\} =\gd \ds\quad &, \quad
\{\dot r(\t,\s)\, ,\, \dot t(\t,\s')\} ={2\gd \over \sh r\, \ch r}\, \dot t
\, \ds\ ,\cr  }
\eqn\tiv$$
and those derived from them by applying $\d_\s^n \d_{\s'}^m$.

Before one can compute the Poisson bracket algebra of the $T, V^\pm$ one
has to rewrite them in terms of the fields and their momenta.  This means
in particular that second (and higher) $\t$-derivatives have to be
eliminated first, using the equations of motion. One might object that one
is not allowed to use the \eoms in a canonical formulation. However, the
conserved quantities given in the previous section are only defined up to
terms that vanish on solutions of the equations of motion. So the correct
starting point for a canonical formulation are the expression where all
higher $\t$-derivatives are eliminated, while the expressions given in
section 2 are merely derived from the canonical ones by use of the
equations of motion.
One has, for example,  using the $\f$-equation of motion \di\
$$\du^2\f={1\over 4}(\ddot\f+2\dot\f'+\f'')={1\over 2}(\f''+\dot\f')+\ch 2r
\, e^{2\f}=\d_\s\du\f+\ch 2r\,  e^{2\f}
\eqn\tv$$
where in the canonical formalism
$$\du r= {1\over 2}(\gd \pr+ r')\ ,\
\du t= {1\over 2}(\gd \th^{-2} r\,  \pt+ t')\ ,\
\du \f= {1\over 2}(\gd \pf+ \f')\ .
\eqn\tva$$
The canonical expressions for the $++$ components of the constraints are
$$\eqalign{
T&=(\du r)^2+\th^2 r\, (\du t)^2 + (\du\f)^2 -(\du\f)' -\ch 2r
e^{2\f}\cr
V^\pm&={1\over \sqrt{2}} e^{\pm i\nu}\Big[ 2\du\f\du r\pm 2i \th r\du\f\du t
+  2\th^3 r(\du t)^2 \mp 2i \th^2 r\du r\du t\cr
&\phantom{=e^{\pm i\nu}\Big[}+{\sh r\over \ch^3 r}\du t\, t'\mp i {\du
r\, t'+\du t\, r'\over \ch^2 r}-(\du r)'\mp i\th r(\du t)'-\sh 2r\,
e^{2\f}\Big]\ \cr }
\eqn\tvb$$
where the substitutions \tva\ are understood.

It is easy to derive from \tiv\ the following Poisson brackets needed for
computing $\{T,T\}$ (I do not write the $\t$-argument any longer; all
Poisson brackets are at equal $\t$.):
$$\eqalign{
\{\du\f(\s)\, ,\, \du\f(\s')\}&={\gd\over 2} \dsp\cr
\{\du\f(\s)\, ,\, \du^2\f(\s')\}&=-{\gd\over 2} \dspp-\gd \ch 2r\,  e^{2\f}
\ds\cr
\{\du^2\f(\s)\, ,\, \du^2\f(\s')\}&=-{\gd\over 2} \dsppp-\gd
(\d_\s-\d_{\s'})\left[ \ch 2r\,  e^{2\f} \ds\right]\cr
\{\du r(\s)\, ,\, \du r(\s')\}&={\gd\over 2} \dsp\cr
\{\du t(\s)\, ,\, \du t(\s')\}&={\gd\over 4}
(\d_\s-\d_{\s'})\left[\th^{-2}r\, \ds \right]\cr
\{\du r(\s)\, ,\, \du t(\s')\}&={\gd\over \ch r\, \sh r}
\left(\du t-{t'\over 2}\right)\ds\ . \cr
}
\eqn\tvi$$
Then one obtains
$$\gmd \{T(\s)\, ,\, T(\s')\}  =
(\d_\s-\d_{\s'})\left[ T(\s') \ds\right]-{1\over 2} \dsppp
\eqn\tvii$$
and similarly
$$\gmd \{\Tb(\s)\, ,\, \Tb(\s')\}  =
-(\d_\s-\d_{\s'})\left[ \Tb(\s') \ds\right]+{1\over 2} \dsppp
\eqn\tviii$$
while
$$\{T(\s)\, ,\, \Tb(\s')\}  =0\ .
\eqn\tix$$
This are just two copies of the conformal algebra. If $\s$ takes values on
the unit circle one can define the modes
$$\eqalign{
L_n&=\gmd \int_{-\pi}^{\pi}\rd\s\left[ T(\t,\s)+{1\over 4}\right]
e^{in(\t+\s)}\cr
\bar L_n&=\gmd \int_{-\pi}^{\pi}\rd\s\left[ \Tb(\t,\s)+{1\over 4}\right]
e^{in(\t-\s)}\ .\cr
}
\eqn\tx$$
Then the brackets \tvii\ and \tviii\ become two Virasoro algebras
$$\eqalign{
i\{L_n\, ,\, L_m\} &=(n-m)L_{n+m}+{c\over 12} (n^3-n)\delta_{n+m,0}\cr
i\{\bar L_n\, ,\, \bar L_m\} &=(n-m)\bar L_{n+m}+{c\over 12}
(n^3-n)\delta_{n+m,0}\cr
}
\eqn\txi$$
while $\{L_n\, ,\, \bar L_m\}=0$. Here $c$ is the central charge given by
$$c={12 \pi\over \gd}\ .
\eqn\txii$$
The occurrence of a central charge already at the classical, Poisson bracket
level is due to the $\d^2\f$-term in $T$. This is reminiscent of the
well-known Liouville theory. The factor $i$ on the left hand side of
equations \txi\ may seem strange at first sight, but one should remember
that  quantization replaces $i$ times the canonical Poisson bracket by the
commutator. Hence \txi\ is indeed the Poisson bracket version of the
(commutator) Virasoro algebra.

In order to compute Poisson brackets involving $V^\pm$ one needs the
Poisson brackets involving the field $\nu$. Now, $\nu$ is only defined
through its partial derivatives, and thus only up to a constant. This
constant, however, may have a non-trivial Poisson bracket with certain modes
of $\nu$ and/or of the other fields. From \div\ one has using \tii
$$\nu'=\gd\pt+t'
\eqn\txiii$$
whereas one does not need $\dot\nu$ explicitly. Equation \txiii\ implies
$$
\d_\s\d_{\s'}\{\nu(\s)\, ,\, \nu(\s')\}=\{\nu'(\s)\, ,\, \nu'(\s')\} = 2\gd
\dsp\ .
\eqn\txiv$$
This is integrated to yield $\{\nu(\s)\, ,\, \nu(\s')\}=-\gd \es +h(\s)-
h(\s')$ where I already used the antisymmetry of the Poisson bracket.
$\es$ is defined to be $+1$ if $\s>\s'$, $-1$ if $\s<\s'$ and $0$ if
$\s=\s'$. The freedom to choose the function $h$ corresponds to the
above-mentioned freedom to add a constant $\nu_0$ to $\nu$ with $\{\nu(\s)\,
,\, \nu_0\}=h(\s)$. However, if one imposes  invariance under translations
$\s\to\s+ a,\ \s'\to\s' +a$ then $h$ can be only linear and one arrives at
$$\{\nu(\s)\, ,\, \nu(\s')\}=-\gd \e_\a(\s-\s')\equiv -\gd\left[
\es-{\a\over\pi} (\s-\s')\right]\ .
\eqn\txv$$
There is only one free parameter $\a$ left. Roughly speaking, $\a$ is
related to the Poisson bracket of the ``center of mass"  position
and momentum (zero) modes of $\nu$. Hence  the choice of $\a$ depends on
the topology of the world-sheet. If $\s$ takes values on the unit circle
$S^1$ with $\s\in[-\pi,\pi]$ and if $\nu$ is defined to be periodic on the
circle, then $\a=1$ so that $\e_1(-\pi)=\e_1(\pi)$ and $\e_1$ can be
continued as a periodic function. If $\s\in {\bf R}$ it is most appropriate
to define $\nu$ such that $\a=0$. For simplicity, I will assume the latter
case till further notice. Later on, I will come back to $\s\in S^1$. The
Poisson brackets one needs thus are
$$\eqalign{
\{\nu(\s)\, ,\, \nu(\s')\}=-\gd \es \quad &, \quad
\{\nu(\s)\, ,\, \du t(\s')\}={\gd\over 2}(1+\th^{-2}\, r) \ds\ , \cr
\{\nu(\s)\, ,\, t'(\s')\}=\gd \ds \quad &, \quad
\{\du t(\s)\, ,\, t'(\s')\}={\gd\over 2}\th^{-2}\, r(\s)\, \dsp\ , \cr
}
\eqn\txvi$$
together with \tvi, and also $\{\nu,r\}=\{\nu,\du r\}=\{\nu,\f\}
=\{\nu,\du\f\}=0$.

Before doing the actual calculation it is helpful to show how the result is
constrained by dimensional and symmetry considerations. Consider first
$\{V^+(\s)\, ,\, V^+(\s')\}$. Each $V^+$ contains a factor
$e^{i\nu}$. The Poisson bracket of $e^{i\nu(\s)}$ with $e^{i\nu(\s')}$
leads to a term $\gd \es V^+(\s) V^+(\s')$. All other terms are
local, i.e. involve $\ds$ or derivatives of $\ds$. On dimensional
grounds\foot{
The naive dimensional counting assigns dimension 1 to each derivative,
dimension 0 to all fields $r,t,\f, \nu$ and functions thereof, except for
functions of $\f$. $e^{2\f}$ has dimension 2 as seen from the action,
while $\ds$ has dimension 1.}
$\ds$ must be multiplied by a dimension 3 object (3 derivatives) and $\dsp$
by a dimension 2 object. Furthermore, these objects must contain an
overall factor $e^{2i\nu}$. If the $T, V^+$ and $V^-$ form a closed
algebra, there are no such objects. The same reasoning applies to
$\{V^-(\s)\, ,\, V^-(\s')\}$. Thus one expects
$$\gmd\{V^\pm(\s)\, ,\, V^\pm(\s')\}=\es
V^\pm(\s)V^\pm(\s')\ .
\eqn\txvii$$
It is a bit tedious, but otherwise straightforward to verify that this is
indeed correct. Note that it is enough to do the computation for $V^+$
since $V^-$ is the complex conjugate of $V^+$ (treating the fields $r,t,\f$
and $\nu$ and their derivatives as real):
$$V^-=(V^+)^*\ .
\eqn\txviii$$

What can one say about $\{V^+(\s)\, ,\, V^-(\s')\}$? Using
\txviii\ one sees that $\{V^+(\s)\, ,\, V^-(\s')\}^* =
-\{V^+(\s)\, ,\, V^-(\s')\}\vert_{\s\leftrightarrow \s'}$. This
fact, together with the same type of arguments as used above, imply
$$\eqalign{
\gmd \{V^+(\s)\, ,\, V^-(\s')\}=&-\es
V^+(\s)V^-(\s') +(\d_\s-\d_{\s'})[a\ds] \cr
&+i b \ds +i(\d_\s^2+\d_{\s'}^2)[d\ds] +\tilde c \dsppp\cr
}
\eqn\txix$$
where $a,b,d,\tilde c$ are real and have (naive) dimensions 2,3,1 and 0.
Hence $\tilde c$ is a c-number. Also, $b,a,d$ cannot contain a factor
$e^{\pm i \nu}$. If one assumes closure of the algebra one must have $b=d=0$
and $a\sim T_{++}$. After a really lengthy computation one indeed finds
equation \txix\ with $b=d=0$, $a=T_{++}$ and $\tilde c=-{1\over 2}$.

Finally the Poisson bracket of $T$ with $V^\pm$ simply shows that
$V^\pm$ are conformally primary fields of weight (conformal dimension)
2. The complete algebra thus is
$$\eqalign{
\gmd \{T(\s)\, ,\, T(\s')\}  &=
(\d_\s-\d_{\s'})\left[ T(\s') \ds\right]-{1\over 2} \dsppp \cr
\gmd \{T(\s)\, ,\, V^\pm\s')\}  &=
(\d_\s-\d_{\s'})\left[ V^\pm(\s') \ds\right]\cr
\gmd \{V^\pm(\s)\, ,\, V^\pm(\s')\}&=\es
V^\pm(\s)V^\pm(\s')\cr
\gmd \{V^\pm(\s)\, ,\, V^\mp(\s')\}&=-\es
V^\pm(\s)V^\mp(\s')\cr
&\phantom{=}+(\d_\s-\d_{\s'})\left[ T(\s') \ds\right] -{1\over 2}
\dsppp \ .\cr }
\eqn\txx$$
The algebra of the $--$ components $\Tb, \Vb^\pm$ is obtained by applying the
star operation, and looks exactly the same except for the replacements $\s\to
-\s,\ \s'\to -\s'$ and hence $\d\to -\d,\ \es\to -\es$.

The algebra  \txx\ is the correct algebra for $\s\in {\bf R}$. If $\s\in
S^1$, as outlined above, one must replace $\es\to \e_1(\s-\s')$ which is a
periodic function. Also $\ds\to {1\over 2}\d_\s \e_1(\s-\s')=\ds-{1\over
2\pi}$ while $\dsp$ remains unchanged. But since the right hand sides of
\txx\ can be written using only $\es,\, \dsp$ and $\dsppp$, only the
replacement $\es\to \e_1(\s-\s')$ is relevant. One then defines the modes
$$\eqalign{
V^\pm_n&= \gmd \int_{-\pi}^\pi \rd\s V^\pm(\t,\s)\, e^{in(\t+\s)}\cr
\bar V^\pm_n&= \gmd \int_{-\pi}^\pi \rd\s \Vb^\pm(\t,\s)\,
e^{in(\t-\s)}\ .\cr
}
\eqn\txxi$$
The mode algebra is\foot{As a further consistency check one can verify that
the Jacobi identities are satisfied.}
$$\eqalign{
i\{L_n\, ,\, L_m\} &=(n-m)L_{n+m}+{c\over 12} (n^3-n)\delta_{n+m,0}\cr
i\{L_n\, ,\, V^\pm_m\} &=(n-m)V^\pm_{n+m}\cr
i\{V^\pm_n\, ,\, V^\pm_m\} &={12\over c}\sum_{k\ne 0} {1\over k}
V^\pm_{n+k}V^\pm_{m-k}\cr
i\{V^\pm_n\, ,\, V^\mp_m\} &=- {12\over c}\sum_{k\ne 0} {1\over k}
V^\pm_{n+k}V^\mp_{m-k}+ (n-m)L_{n+m}+{c\over 12} (n^3-n)\delta_{n+m,0}\ .\cr
}
\eqn\txxii$$
This is a non-linear algebra, reminiscent of the $W$-algebras. What is new
here are the non-local terms involving $\es$, or the ${1\over k}$ in the
mode algebra.

\section{The Poisson brackets of the $f_i$ and $g_i$}

In this subsection, I will deduce the Poisson brackets of the $f_i$ from
those of $T, V^+$ and $V^-$. Those of the $g_i$ follow by
the star operation.  As in the previous section, the
$T, V^\pm$ as well as the $f_i$ are considered as functions on
phase-space  without imposing the equations of motion, so they are supposed
to depend on $\t$ and $\s$. Again, since all Poisson brackets are at equal
$\t$, I will not write $\t$ explicitly. Recall
$$\eqalign{
T&={1\over 4}(f_1')^2-{1\over 2}f_1''+f_+'(f_-'-f_-^2f_+')\cr
V^\pm&={1\over \sqrt{2}}(f_1'-\d_\s) O^\pm\cr}
\eqn\txxiii$$
where now $f_i'=\d_\s f_i$, and where I introduced
$$\eqalign{
O^+&=\exp\left( -2\int^\s B\right) (f_-'-f_-^2f_+')\cr
O^-&=\exp\left(  2\int^\s B\right) f_+' \cr
B&=f_-f_+'\ .\cr
}
\eqn\txxiv$$
One observes that the part of $T$ involving only $f_1$ already satisfies
the correct Virasoro Poisson bracket algebra with central charge $c$ by
itself if
$$\{ f_1(\s)\, ,\, f_1(\s')\}=2\gd\dsp\ .
\eqn\txxv$$
It is thus natural to use this and
$$\{ f_1(\s)\, ,\, f_\pm(\s')\}=0
\eqn\txxv$$
as an Ansatz.

Consider now  $\{V^-(\s)\, ,\, V^-(\s')\}=\gd \es
V^-(\s) V^-(\s')$. By the same type of argument as used in the previous
subsection one must have
$$\{O^-(\s)\, ,\, O^-(\s')\}=[\a(\s)\b(\s')+\a(\s')\b(\s)]\es
+(\d_\s-\d_{\s'})[\xi(\s')\ds ]
\eqn\txxvii$$
where $\a$ and $\b$ are functionals of $f_\pm$ and  have (naive) dimension 1
and $\xi$ (naive) dimension 0. This leads to a term $\sim \dsppp$ in
$\{V^-(\s)\, ,\, V^-(\s')\}$ unless $\xi=0$. It is also easily seen that in
order to reproduce  $\{V^-(\s)\, ,\, V^-(\s')\}$ correctly one must identify
$\a=\b={\g\over \sqrt{2}} O^-$, hence
$$\{O^-(\s)\, ,\, O^-(\s')\}=\gd \es O^-(\s) O^-(\s') \ .
\eqn\txxviii$$
This looks identical to the bracket of $V^-$ with itself,  but
$O^-=e^{2\int^\s B}f_+' $ is a simpler object than $V^-$. One has
$$\eqalign{
\{e^{2\int^\s B}f_+'(\s)\, ,\, e^{2\int^{\s'} B}f_+'(\s')\}
=&e^{2\int^\s B}e^{2\int^{\s'} B}\Big[
\{f_+'(\s)\, ,\, f_+'(\s')\}\cr
&+2 f_+'(\s) \{\int^\s B \, ,\, f_+'(\s')\}
+2 \{f_+'(\s)\, ,\, \int^{\s'} B \}f_+'(\s')\cr
&+4  f_+'(\s)\{\int^\s B \, ,\, \int^{\s'} B \}f_+'(\s') \Big] \ .
}
\eqn\txxix$$
A natural Ansatz is
$$\eqalign{
\{\int^\s B \, ,\, \int^{\s'} B \}&={a\over 4}\gd\es \cr
\{\int^\s B \, ,\, f_+'(\s')\}&={b\over 4}\gd f_+'(\s') \es \cr
\{f_+'(\s)\, ,\, f_+'(\s')\}&=d \gd f_+'(\s) f_+'(\s') \es\ . \cr
}
\eqn\txxx$$
Then \txxviii\ is satisfied if $a+b+d=1$. Taking derivatives of the Poisson
brackets \txxx\ one obtains the brackets of $B\equiv f_-f_+'$ with itself
and with $f_+'$. From these one then also deduces the Poisson brackets of
$f_-$ with itself and with $f_+'$. Let me anticipate that $a$ must vanish.
With $a=0$ one has
$$\eqalign{
\{f_-(\s)\, ,\, f_-(\s')\}&=d \gd f_-(\s) f_-(\s') \es \cr
\{f_-(\s)\, ,\, f_+'(\s')\}&=-d \gd f_-(\s) f_+'(\s') \es
+{b\over 2} \gd \ds \cr
\{f_+'(\s)\, ,\, f_+'(\s')\}&=d \gd f_+'(\s) f_+'(\s') \es \ .\cr }
\eqn\txxxi$$
{}From these Poisson brackets it follows
$$\eqalign{
\{(f_-'-f_-^2f_+')(\s)\, ,\, (f_-'-f_-^2f_+')(\s')\}&=
d \gd (f_-'-f_-^2f_+')(\s) (f_-'-f_-^2f_+')(\s') \es \cr
&\phantom{=}
-\left({b\over 2}+d\right)\gd (\d_\s-\d_{\s'}) [f_-^2(\s')\ds ] \cr
\{B(\s)\, ,\, (f_-'-f_-^2f_+')(\s')\}&=
{b\over 2}\gd f_-(\s') \dsp\cr
&\phantom{=} -{b\over 2}\gd (f_-'-f_-^2f_+')(\s')\ds\ . \cr }
\eqn\txxxii$$

Exactly as eq. \txxviii\ was derived, one must have the same Poisson
bracket with $O^-\to O^+$. Using the definition \txxiv\ of $O^+$ and the
Poison brackets \txxxii\ (recalling that $\{B,B\}\sim a=0$) one finds
$$\eqalign{\{O^+(\s)\, ,\, O^+(\s')\}=&(d+b)\gd \es O^+(\s) O^+(\s') \cr
&- \left( {b\over 2}+d\right) \gd (\d_\s-\d_{\s'})
\left[ e^{-4\int^{\s'} B} f_-^2(\s') \ds \right] \ .}
\eqn\txxxiii$$
Had $a$ not been zero,
a term $\sim \dsppp/(f_+'(\s) f_+'(\s'))$ would have appeared, among others,
on the right hand side.
Hence one must choose $a=0,\ {b\over 2}+d=0,\ d+b=1$, or
$$d=-1\ ,\quad b=2\ , \quad a=0\ .
\eqn\txxxiv$$
It is then straightforward to verify that the Poisson brackets
$$\eqalign{
\{f_1(\s)\, ,\, f_1(\s')\}&=2\gd \dsp\cr
\{f_1(\s)\, ,\, f_\pm(\s')\}&=0\cr
\{f_-(\s)\, ,\, f_-(\s')\}&=- \gd f_-(\s) f_-(\s') \es \cr
\{f_+'(\s)\, ,\, f_+'(\s')\}&=- \gd f_+'(\s) f_+'(\s') \es \cr
\{f_-(\s)\, ,\, f_+'(\s')\}&= \gd f_-(\s) f_+'(\s') \es
+\gd \ds \cr }
\eqn\txxxv$$
do imply {\it all} six Poisson brackets \txx\ of the $T$, $V^+$ and
$V^-$.\foot{
In particular, it is easy to obtain $\{V^+,V^-\}$ using these $f$ Poisson
brackets  while it was not at all easy using the canonical $r,t,\f$ and
$\nu$ Poisson brackets.} Note that I have not completely proven the
converse, i.e. I have not {\it proven} that \txxxv\ is the only possible
choice of $f_i$ Poisson brackets that implies the $T, V^\pm$ Poisson
brackets. On the other hand it is hard to see how this could fail to be
true. Indeed, I deduced \txxxv\ from only two Poisson brackets
$\{V^\pm,V^\pm\}$ and a simple Ansatz. The fact that all other four Poisson
brackets $\{T,T\}$, $\{T, V^\pm\}$ and $\{V^+,V^-\}$ then are  satisfied,
very strongly supports the conjecture, that \txxxv\ is indeed the only
possible Poisson bracket of the $f_i$. Obviously, the same Poisson brackets
are valid if one replaces $f_i\to g_i$ (star operation).

It also follows that
$$\eqalign{
\{T(\s)\, ,\, f_\pm(\s')\}&=-f_\pm(\s') \ds \cr
\{T(\s)\, ,\, e^{f_1(\s')}\}&=-\d_{\s'}\left[ e^{f_1(\s')} \ds \right]\cr
}
\eqn\txxxvi$$
so that $f_+$ and $f_-$ are conformal primaries with conformal dimension
$0$ and $e^{f_1} $ is primary with dimension 1. This is indeed needed to
ensure that the functions $F_i$ defined in section 2 have  well-defined
conformal properties: they are all conformal primaries of dimension zero,
i.e. scalar functions. It then also follows that the ``building blocks" for
the classical solutions $V,W,X,Y,Z$ are conformal primaries with dimensions
zero with respect to $T_{++}$ and $T_{--}$. Moreover, from \dxxxxx\ and
\dxxxxxiii\ one sees that $r$ and $t$ are  also dimension zero primaries
(scalars) while $e^{2\f}$ has conformal dimension $(1,1)$ (i.e. is a
scalar density) as it should in order that the action and equations of
motion be conformally invariant.

\section{Transformation to free fields (harmonic oscillators)}

While the Poisson bracket of $f_1$ with itself  is a free field bracket,
i.e. its Fourier modes $(f_1)_n$ have harmonic oscillator Poisson brackets,
this is not true for the brackets \txxxv\ of $f_+,\, f_-$. The bracket of
$f_-$ with itself implies
$$\{\d_\s \log f_-(\s)\, ,\, \d_{\s'} \log f_-(\s')\}=2\gd \dsp
\eqn\txxxvii$$
which is a free field Poisson bracket, just as for $f_1$. The same is true
if one replaces $f_-$ by $f_+'$. However, $f_-$ has conformal dimension
zero while $f_+'$ has conformal dimension one. Also the Poisson bracket of
$f_-$ with $f_+'$ is non-vanishing. It is then natural to set
$$\eqalign{
f_-&=e^{\sqrt{2}\vf_1}\cr
f_+'&={1\over \sqrt{2}}e^{-\sqrt{2}\vf_1} (\d_\s\vf_1+i\d_\s\vf_2)\cr
f_1&=\sqrt{2}\vf_3\cr
}
\eqn\txxxviii$$
A straightforward computation then shows that the Poisson brackets \txxxv\
are equivalent\foot{
Of course, $\{\vf_i',\vf_j'\}$ follows from $\{\vf_i,\vf_j\}$ while the
converse is only true up to a function of integration. However as already
discussed earlier (see the discussion of $\{\nu,\nu\}$ in section 3.1) this
ambiguity is only related to the zero-modes of $\vf_j$ and can be fixed in
a natural way, thus replacing e.g. $\es\to\e_1(\s-\s')$ if $\s,\s'\in S^1$.}
to
$$\{\d_\s \vf_i(\s)\, ,\, \d_{\s'} \vf_j(\s')\}=\gd \delta_{ij} \dsp
\eqn\txxxix$$
or
$$\{\vf_i(\s)\, ,\, \vf_j(\s')\}=-{\gd\over 2} \delta_{ij} \es\ .
\eqn\txxxx$$
If one considers $\s\in S^1$, the mode expansion\foot{
Note that this is a Fourier expansion in $\s$. The factor $e^{-in\t}$ is
only extracted from the $\vf_n^j$ for convenience. The $\vf_n^j$ are still
functions of $\t$. It is only if one imposes the equations of motion that
the $\vf_n^j$ are constant.}
$$\d_\s \vf_j(\t,\s)={\g\over \sqrt{2\pi}} \sum_n
\vf^j_n e^{-in(\t+\s)} \eqn\txxxxi$$ %
leads to
$$i\{\vf^j_n\, ,\, \vf^k_m\}=n \delta^{ij} \delta_{n+m,0}
\eqn\txxxxii$$
as appropriate for three sets of harmonic oscillators.

The conserved quantities $T, V^\pm$ are easily expressed in terms of the
$\vf_j$ as
$$\eqalign{
T&={1\over 2} \sum_{j=1}^3 (\d_\s \vf_j)^2 -{1\over \sqrt{2}}\d_\s^2\vf_3\cr
V^\pm&={1\over 2}(\sqrt{2}\d_\s\vf_3-\d_\s)\left[ e^{\mp i\sqrt{2}\vf_2}
(\d_\s\vf_1\mp i\d_\s\vf_2)\right]\ .\cr
}
\eqn\txxxxiii$$
Thus in terms of the $\vf_j$ the $V^\pm$ are {\it local} expressions of the
fields, analogous to standard vertex operators. (Of course, their Poisson
brackets exhibit the non-local $\es$-function.)

Again, the same formulas apply if one replaces $f_j\to g_j$, $\vf_j\to
\bar\vf_j$ and $T=T_{++}, V^\pm=V^\pm_{++}\to T_{--}, V^\pm_{--}$ (and
$i\to -i$).

\chapter{Conclusions and open problems}

In this final section, after summarizing of what has been done, I will
briefly discuss a couple of issues which should be addressed but for which
the solution has yet to be worked out. These are:
periodicity in $\s$ and zero-modes, periodicity in the target space,
the associated linear differential equation and the hierarchy of non-linear
integrable partial differential equations, the screening fields and the
whole issue of quantization, in particular the structure of the quantum
constraint algebra, and last, but not least, the relevance to black hole
physics.

\section{Conclusions}

In this article, I have considered a $1+1$ dimensional field theory that
describes a string propagating on a two-dimensional (euclidean) black hole
background. This string also has an ``internal" degree of freedom, which is
coupled through a non-trivial tachyon potential to the radial target-space
field. Alternatively one can view this as a string propagating on a
three-dimensional background which is the product of a two-dimensional
blackhole and  a flat direction. I have investigated the corresponding
{\it classical} theory. Using the formal solution of refs. \LS\ and \GS, I
have given the general explicit solution to the equations of motion. It
expresses the interacting fields through a rather involved
transformation in terms of functions $f_i(u)$ and $g_i(v)$ of one
light-cone variable only. The theory has three left-moving and three
right-moving conserved quantities. The right (left)-moving
conserved quantities form a closed non-linear, non-local Poisson bracket
algebra. This algebra is a Virasoro algebra extended by two
conformal dimension-two primaries. The name ``$V$-algebra" seems
appropriate, stressing the similarities (non-linearity) and differences
(non-locality) with the well known $W$-algebras (see e.g.
\REF\AB{A. Bilal, Introduction to W-algebras, in: Proc. Trieste Spring
School on String Theory and Quantum Gravity, 1991, J. Harvey et al eds.,
World Scientific.}
ref. \AB) .
{}From these Poisson brackets I obtained the Poisson brackets
of the $f_i$ and $g_i$, and finally, after an ultimate transformation those
of $\vf_i$ and $\bar\vf_i$. The $\vf_i$ ($\bar\vf_i$) are left (right)-moving
free fields with harmonic oscillator Poisson brackets.

\section{Periodicity in $\s$ and zero-modes}

If one considers the theory as given by \uiii, \di\ as a $1+1$ dimensional
field theory on the infinite real line, i.e. $\s\in {\bf R}$, then nothing
special has to be observed and in particular the $\es$-function is just the
usual one ($=+1$ for $\s>\s'$ and $=-1$ if $\s'<\s$). If however, one
thinks of the theory as a theory of closed strings, then $\s\in S^1$ and
one must have periodicity in $\s$ of all {\it physical} quantities. For
example, the field $\nu$ itself is not physical, and hence need not be
periodic. On the other hand, the conserved quantities (constraints) $T$ and
$V^\pm$  are physical and should be periodic. Hence $e^{\pm i\nu}$ and
$\du\nu, \dv\nu$ must be periodic, and under $\s\to\s+2\pi$, one must have
$\nu\to\nu+2\pi n,\ n\in {\bf Z}$. Moreover, the fiels $r,\, t,\, \f$
certainly are physical and should be periodic under $\s\to\s+2\pi$. The
issue of periodicity is probably best discussed when the fields are
expressed in terms of the $\vf_j$ and $\bar\vf_j$. The non-zero modes of
$\vf_j$ and $\bar\vf_j$ are automatically periodic, while the zero-mode
piece $(\vf_j)_0=q_j+p_j {\t+\s\over 2\pi}$ gives $\vf_j\to \vf_j+p_j$, and
similarly $\bar\vf_j\to \bar\vf_j+\bar p_j$. One can then work out how the
fields $r,t$ and $\f$ change under $\s\to\s+2\pi$, and what conditions have
to be imposed on the $p_j, \bar p_j$ to make $r, t, \f$ periodic.\foot{
Actually it does not seem to be a priori garantied that one can achieve
periodicity of $r, t, \f$. If one could not, then all the developments in
this paper would only make sense as a field theory with $\s\in {\bf R}$.
However, this would be most surprising since one expects a string theory to
make sense on the black hole background, at least classically.}

\section{Periodicity in the target space}

Another related issue is periodicity in the target space. Given that the
action \uiii\ describes a $\s$-model for a string on the euclidean black
hole manifold, one should have $r\ge 0$ and $t$ should have period $2\pi$,
i.e. $t$ and $t+2\pi$ should be identified. $r\ge 0$ can be easily achieved
by choosing the appropriate branch of the square root of $\sh^2 r$.
Identifying $t$ with $t+2\pi$ allows for more freedom in the zero-mode
discussion of the previous subsection. One can (and has to) include winding
modes: $t\to t+2\pi m$ as $\s\to \s+2\pi$ corresponds to a string winding
$m$ times around the semi-infinite cigar-shaped 2D black hole manifold.

\section{Associated linear differential equation and hierarchy of integrable
non-linear  partial differential equations}

{}From experience with integrable models, in particular the (non-affine,
conformally invariant) Toda models
\REF\BG{A. Bilal and J.-L. Gervais, \PL B206 1988 412 ;
\NP B314 1989 646 ,
{\bf B318} (1989) 579 .}
[\BG] one expects that the conserved quantities appear as coefficients of
an ordinary linear differential equation, e.g. for the $A_{n-1}$ Toda model
$$\left[\du^n-\sum_{k=2}^n W^{(k)}(u) \du^{n-k}\right]\psi(u) =0\ .
\eqn\qi$$
The $ W^{(k)}(u)$ are the conserved quantities which form the
$W_n$-algebra, while the solutions $\psi_j(u)$ of this equation, together
with the solutions $\chi_j(v)$ of a similar equation in $v$, are the
building blocks of the general solution to the Toda equations of motion.
The $ W^{(k)}(u)$ have (naive) dimension $k$.

In the present theory all conserved quantities have dimension 2. The theory
being non-abelian in character, one may guess a linear differential
equation of the form
$$\left[ \d_u^2-\pmatrix{ \a T(u)& \b_+ V^+(u)\cr \b_- V^-(u) & \delta T(u)
\cr } \right] \Psi(u)=0\ .
\eqn\qii$$
It should be stressed that, at present, this should only be regarded as a
{\it guess} about the {\it form} of the fundamental differential equation. If
\qii\ or some similar differential equation is true, one may explore the
whole hierarchy of  non-linear  partial differential equations associated
with it, the same way as the KdV hierarchy is associated with
$(\d^2-T)\psi=0$ or the KP hierarchy is associated with \qi.

Actually, one can at least find one differential equation of type \qii\
and its solution. From experience with Toda theories one can try a simple
Ansatz:
$$\eqalign{
\psi_1&=\exp(a\vf_1 +ib\vf_2 +d\vf_3)\cr
\psi_2&=\exp(a\vf_1 -ib\vf_2 +d\vf_3)\ .\cr
}
\eqn\qqi$$
Then using the form \txxxxiii\ of $T$ and $V^\pm$ one finds that the above
differential equation \qii\ is satisfied if and only if
$$\eqalign{
a={1\over \sqrt{2}}\quad &, \quad b=d= -{1\over \sqrt{2}}\cr
\a=\delta=1 \quad &, \b_+=\b_-=-\sqrt{2}\ .\cr
}
\eqn\qqii$$

\section{Screening fields and quantization}

A natural starting point for quantization are the Poisson brackets \txxxx\
of the free fields $\vf_j, \bar\vf_j$ or, if $\s\in S^1$ of their Fourier
modes \txxxxii. Let's suppose $\s\in S^1$. Quantization then gives
$$\eqalign{
[\vf_n^j\, ,\, \vf_m^k]&= [\bar\vf_n^j\, ,\, \bar\vf_m^k]=n\delta^{jk}
\delta_{n+m,0} \cr
[\vf_n^j\, ,\, \bar\vf_m^k]&=0\cr
}
\eqn\qiii$$
together with the appropriate commutators with possible zero-modes $q^j, \bar
q^j$. Before bothering about zero-modes one can already write down the
stress tensor:
$$T(\s)={1\over 2}\sum_{j=1}^3 :(\d_\s \vf_j)^2: -{1\over
\sqrt{2}}\d_\s^2\vf_3
\eqn\qiv$$
(which is normalized as in section 3, and thus differs from the usual
normalization by a factor of $\gd/(2\pi)$) or the (correctly normalized)
Virasoro generators (recall \tx\ and \txxxxi)
$$\eqalign{
L_n&={1\over 2}\sum_m \sum_{j=1}^3 :\vf^j_m \vf^j_{n-m}: +\a_0 n \vf_n^3
-{\a_0^2\over 2}\delta_{n,0}\cr
\a_0&=i{\sqrt{\pi}\over \g}\cr
}
\eqn\qv$$
which is of the standard\foot{
Usually one redefines $a_n^j=\vf_n^j-\a_0 \delta_{n,0}\delta^{j,3}$ so that
the Virasoro generators take the more familiar form $L_n={1\over 2}\sum_m
\sum_{j=1}^3 :a^j_m a^j_{n-m}: +\a_0 (n+1) a_n^3$.} Feigin-Fuchs form with
background charge $\a_0$. Without further computation one knows that the
$L_n$ satisfy a Virasoro algebra with quantum central charge given by
$$c=3-12\a_0^2=3+{12\pi\over \gd}
\eqn\qvi$$
where the contribution $-12\a_0^2={12\pi\over \gd}$ was already present at
the classical level (cf. \txii) while the $3$ is the quantum contribution
of the three fields.

In order to be able to construct a quantum version of the classical
transformations from the $\vf^j, \bar\vf^j$ to the fields  $f_j, g_j$ and
then to $r, t, \f$ one needs to ensure, among other things, that the
integrals defining the $F_i, G_i$ (eqs. \dxxxvi-\dxxxviii) have conformal
dimension zero, i.e. are screening operators. One possibility could be to
replace
$$e^{f_1}\equiv e^{\sqrt{2}\vf_3}\quad \to \quad :e^{\sqrt{2}\a\vf_3}:
\quad {\rm with} \quad \a-{\gd\over 2\pi} \a^2=1
\eqn\qvii$$
so that it has conformal dimension one, and to modify $f_\pm$ so that they
keep conformal dimension zero and commute with each other and with
$:e^{\sqrt{2}\a\vf_3}:$. Whether this or a similar possibility can be
successfully implemented has still to be worked out.

\section{Structure of the quantum constraint algebra}

As just discussed, upon quantization, the $L_n$ will satisfy the usual
(quantum) Virasoro algebra, obtained from the classical Poisson bracket
relation \txxii\ by the simple substitution $i\{,\}\to [,]$, only the value
of the central charge $c$ is changed with respect to its classical value by a
quantum contribution. What about the other commutators corresponding to
\txxii? Clearly, one expects $V^\pm$ to remain a conformal dimension 2
primary operator so that
$$[L_n\, ,\, V^\pm_m]=(n-m)V^\pm_{n+m}\ .
\eqn\qviii$$
The commutators of $V^\pm$ with $V^\pm$ and $V^\mp$ are expected to contain
a similar bilinear term as in \txxii. However, one has to introduce some
normal-ordering, e.g. with respect to the mode index of the $V^\pm_k$. This
has to be consistent with the Jacobi identities. Note that for the linear
part of the algebra, the Jacobi identities are always satisfied if they are
for the corresponding Poisson brackets. For the non-linear terms however,
the proof of the Jacobi identities for the Poisson brackets \txxii\ used
the fact that $V_{n+k}$ and $V_{m-k}$ on the right hand side commute. This
is no longer true at the quantum level where the ordering is important.
For this reason, it is not clear to me at present what should be the
correct quantum commutator replacing $i\{V^\pm,V^\pm\}$ and
$i\{V^\pm,V^\mp\}$.

\section{Relevance to black hole physics}

One can consider all the developments in this paper merely from the point
of view of integrable  models and conformal field theory. In particular, I
believe that the new $V$-algebra \txxii\ and the possible hierarchy
associated with the linear differential equation \qii\ may lead to exciting
developments.

On the other hand, one may ask whether anything has been gained for our
understanding of strings propagating on the black hole background. At this
point it is too early to give a definite answer. Clearly, first one has to
solve the above-mentioned issue of periodicity and winding-modes. This does
not seem too difficult. Then one should actually quantize the theory. This
may be quite non-trivial, but the rigid structure of integrable models
probably will prove helpful. In particular, the spectrum is expected to
fall into representations of a possible quantum version of the $V$-algebra
\txxii. This in turn gives information about the zero-modes, making it
eventually possible to compute the one-loop partition function (and maybe
the string-theoretic one-loop entropy). Quantization of the free fields
$\vf_j$ is straightforward (and exact to all orders in $\hbar\sim \gd\sim
\a'$ where $\a'$ is the usual inverse string tension). When formulating the
quantum version of the transformation to the original fields $r, t$ and
$f$, one might be forced into a semiclassical expansion in $\hbar\sim\a'$
which can be compared to the well-known $\b$-function equations of ref. \CP.
If however, one is able to formulate this transformation exactly one would
have succeeded in providing  exact, non-perturbative\foot{
Here the term ``non-perturbative" refers to the $\a'$ expansion. The string
world sheet still has cylindrical (or spherical) topology.}
information about a string on the black hole background. It would be most
interesting to compare these results to those of
\REF\DVV{R. Dijkgraaf, H. Verlinde and E. Verlinde, \NP B371 1992 269 .}
ref. \DVV\ for the $Sl(2,{\bf R})/U(1)$ black hole.

\ack

It is a pleasure to acknowledge discussions with Curtis Callan, Jean-Loup
Gervais, David Gross and Ed Witten.

\Appendix{A}
%\centerline{The generators of the Lie algebra $B_2$}

The Lie algebra $B_2$ has two Cartan generators, $h_1$ and $h_2$ and eight
step operators $E_{\pm e_1}, E_{\pm e_2}, E_{\pm e_1\pm e_2}, E_{\pm e_1
\mp e_2}$. Section 2 makes use of the following realization [\GS] of the
generators in terms of five fermionic oscillators $b_0, b_{\pm 1}, b_{\pm
2}$ with $[ b_j\, ,\, b_k^+]_+=\delta_{jk}$.
$$\eqalign{
h_1&=b_1^+b_1-b_{-1}^+b_{-1}-b_2^+b_2+b_{-2}^+b_{-2}\cr
h_2&=2(b_2^+b_2-b_{-2}^+b_{-2})\cr
E_{e_1}&=\sqrt{2}(b_1^+b_0-b_0^+b_{-1})\cr
E_{e_2}&=\sqrt{2}(b_2^+b_0-b_0^+b_{-2})\cr
E_{e_1+e_2}&=b_1^+b_{-2}-b_2^+b_{-1}\cr
E_{e_1-e_2}&=b_1^+b_{2}-b_{-2}^+b_{-1}\cr}
\eqn\ai$$
together with
$$ E_{-\a}=E_\a^+\ .
\eqn\aii$$
The grading element $H$ is given by
$$H=2h_1+h_2=2(b_1^+b_1-b_{-1}^+b_{-1})
\eqn\aiii$$
and simply counts (twice) the oscillators of type $1$ minus those of type
$-1$.

\Appendix{B}
%\centerline{Formulas for the explicit solution}

In this appendix, I collect some useful relations between the functions
$V,W,X,Y,Z$ and their derivatives used in section 2.

{}From
$$\eqalign{
W&=-g_++F_1G_--F_2G_+-f_-X\cr
Y&=1+g_+g_-+F_1(G_1-g_-G_-)+F_2(G_2+g_-G_+)-f_-V]\cr }
\eqn\bb$$
one gets after some algebra
$$W\dv V-X\dv Y=-{1\over 2}\dv Z -g_-g_+'Z
\eqn\bi$$
and using the star operation also
$$X\du Y-V\du W={1\over 2} \du Z+f_-f_+'Z\ .
\eqn\bii$$
Another relation needed to prove the $\au$ (or $\f$) \eom is
$$Z\du\dv Z-\du Z\dv Z-2e^{f_1+g_1} Z=-4e^{f_1+g_1}XY
\eqn\biii$$
which is verified by direct computation, recalling that $\du$ only acts on
$F$'s and $f$'s while $\dv$ only acts on $G$'s and $g$'s, and that $\du
F_1=-e^{f_1}(1+2f_+f_-)$ etc. as seen from the definitions \dxxxvi.

When computing $T_{--}$ the following relation are needed
$$
X\dv W-W\dv X=-g_+' Z+\a
\eqn\bv$$
$$
Y\dv V-V\dv Y=-(g_-'-g_-^2g_+')Z+\b
\eqn\bvi$$
$$
\a\b-{1\over 4}(\dv Z)^2=-(F_1^2+F_2F_3)e^{2g_1}Z
\eqn\bvii$$
$$
-{1\over 2}g_1' \dv Z+{1\over 2}\dv^2 Z-g_+'\b-(g_-'-g_-^2g_+')\a =
{(\dv Z)^2\over 4Z}-{\a\b\over Z}
\eqn\bviii$$
where
$$\eqalign{
\a&=e^{g_1} \left[ 2g_+F_1-F_2+g_+^2F_3-(F_1^2+F_2F_3)(G_--g_+G_+)\right]
\cr
\b&=e^{g_1}\Big\{ 2g_-(1+g_+g_-)F_1-g_-^2F_2+(1+g_+g_-)^2F_3\cr
&\phantom{=e^{g_1}\Big\{   }+(F_1^2+F_2F_3) \left[ (1+g_+g_-)(G_2+g_-G_+)
+g_-(G_1-g_-G_-)\right] \Big\}\ .\cr
}
\eqn\bix$$
Relations \bv\ and \bvi\ are derived much as \bi\ using \bb. Equation
\bvii\ is verified by straightforward algebra, as is also \bviii\ (using
\bvii). When evaluating $V^+_{++}$ one needs
$$(\du-f_1'-2f_-f_+')\b\st +(f_-'-f_-^2f_+')\du Z=0
\eqn\bx$$
which is also not too difficult to be checked.

\refout

\end